\documentclass[pdftex]{ws-ijgmmp}
\usepackage[square,numbers,sort&compress]{natbib}
\RequirePackage{graphicx}
\RequirePackage{mathptmx}      
\RequirePackage{flushend}
\RequirePackage[square,numbers,sort&compress]{natbib}
\RequirePackage[colorlinks,citecolor=blue,urlcolor=blue,linkcolor=blue]{hyperref}

\usepackage{graphicx}	
\usepackage{amsmath}	
\usepackage{amssymb}	
\newcommand{\G}{\mathcal{G}}
\newcommand{\frg}{$F(R,\G) ~$}
\newcommand{\beq}{\begin{equation}}
\newcommand{\eeq}{\end{equation}}

\title{Stability of Hyperbolic and Matter-Dominated Bounce Cosmologies From F(R,G) Modified Gravity at Late Evolution Stages}

\author{
G. Nav\'o, E. Elizalde
\\
Institute of Space Sciences (IEEC-CSIC) Campus UAB, Carrer de Can Magrans, s/n 08193 Barcelona. \\
Recieved 5 December 2019.\\ 
Accepted 24 July 2020.\\
DOI:10.1142/S0219887820501625\\
}

\date{\today}

\begin{document}
\maketitle


%
\catchline{}{}{}{}{}
%

\begin{abstract}
The stability of two different bounce scenarios from \frg modified gravity at later times is studied, namely a hyperbolic cosine bounce model and a matter-dominated one. After describing the main characteristics of \frg modified gravity, the two different bounce scenarios stemming from this theory are reconstructed and their stability at late stages is discussed. The stability of the hyperbolic cosine model is proven, while the  concrete matter-bounce model here chosen does not seem to accomplish the necessary conditions to be stable at later times.

\end{abstract}


\section{Introduction}
\label{sec: intro}

The inflationary scenario (for an extensive review see \cite{inflation_review} and references therein) provides the most popular scheme to answer the main problems of the original Big Bang cosmological model. It is able to explain (with the addition sometimes of quantum correction terms to the classical Einsteinian gravity),  why the current observations converge on a flat, homogeneous and isotropic universe in accelerated expansion. This scenario is characterized by a huge expansion of the universe between the origin and the radiation era and provides the mechanism to generate nearly scale invariant primordial density fluctuations, which are thought to be the seeds of the large structures of our universe. Even though astronomical observations support such primordial fluctuations \cite{inflation_obs_planck}, they do not necessary imply inflation since there are several models in bouncing cosmologies that may also lead to them \cite{branderberger2,odintsov_matter_loop}.

The bounce scenario is actually the most usual alternative to inflation \cite{brandenberger1,branderberger2}. One of its main features is the avoidance of the initial singularity, although there exist bounce models which do present it \cite{bounce_sing3,bounce_sing1,bounce_sing2}. It is characterized by a contraction phase followed by the bounce point, which leads then to an expanding era. The inflection point where the Universe passes from one stage through to the other, the bounce point, occurs at time $t=t_b$. At this time, the scale factor $a(t_b)\equiv a_b$ and the Hubble parameter $H(t_b)\equiv H_b$ are 
\beq
\label{eq: conditions bounce}
\dot{a}_b=0~~,~~\ddot{a}_b\geq0~~;~~ H_b=\frac{\dot{a}_b}{a_b}.
\eeq
Thus, any Hubble rate involving a bounce scenario must fulfill these conditions.

In general, bouncing cosmologies emerge from the idea of avoiding the initial singularity in the scope of the Loop Quantum Cosmology matter bounce (see \cite{status_quantum} and references therein), in  scalar field theories, as for example in \cite{example_scalar1} or  \cite{scalar_2}, and in modified gravities, which offer consistent descriptions of a bounce scenario (as an illustration one can consider \cite{superbounce1,superbounce2} or \cite{lambda_bounce_no_lambda}). In the realms of General Relativity (GR) it is very difficult to handle the bouncing behavior since it leads to the violation of the Null Energy Condition (NEC) caused by the change in the sign of the Hubble rate at the bounce point \cite{nec_review}. An extended discussion on this issue is performed in \cite{NEC_1} and \cite{NEC_2}. In the present work, we focus on \frg modified gravity, which was introduced in \cite{nojiri,cognola,laurentis} as a possible gravitational origin for Dark Energy. In the same vein, the literature provides numerous modified gravities \citep{nojiri_odintsov} which have been explored in the scope of bouncing cosmologies such as $F(R)$ gravity \cite{barragan}, teleparallel gravity \cite{cai}, modified Gauss-Bonnet gravity \cite{bamba}, or loop quantum gravity \cite{olmo}.

A well-known bounce scenario is the matter-dominated one, where the contraction is controlled by matter and, in addition, it is able to generate primordial density perturbations with a nearly scale invariant and adiabatic spectrum compatible with observations \cite{novello,quasimater,haro}. They are produced in the following way. Previous to the matter contraction, the fluctuations are inside the Hubble horizon, considered to be very large at that moment, what allows a vast region of the universe to be at causal contact. Once contraction has started, at some point perturbations become larger than the Hubble horizon, getting out of it and therefore getting frozen. After the bounce, the horizon grows, which makes the fluctuations to enter it again, allowing the production of the observed spectrum. However, this scenario presents some inconveniences to be addressed \cite{brandeburger_progres}. To mention one of the most important,  it arises within the matter-dominated contraction epoch, which can be problematic leading to a growth of anisotropies, the so called BKL-instabilities \cite{bkl_first}. In this work we will not focus on this aspect, conveniently discussed in the literature \cite{bkl_issued}, where the problem is solved by adding an ekpyrotic contraction phase (for a review of the ekpyrotic model, see e.g. \cite{ekpyrotic}).

Another popular bounce model is the hyperbolic cosine bounce. It is quite popular in the literature, since on top of being geodesically complete, it preserves causality and addresses the horizon problem \cite{hyperbolic_bounce3,hyperbolic_bounce,hyperbolic_bounce2}. Furthermore, working with this model one is able to provide relatively elegant solutions.

In this paper we  study the stability at late time of a hyperbolic cosine bounce model and of a matter-dominated bounce scenario from an \frg gravity. We  first provide a brief description of the \frg modified gravity theory,  introducing the formalism and  presenting some illustrative examples of the reconstructing method used. Then, we  reconstruct both the hyperbolic and the matter-dominated models. Finally, we  discuss the stability of the models at the later stages of the universe evolution.

\section{F(R,$\G$) gravity}
\label{sec:grav}
\frg gravity is a type of modified gravity that involves both the Ricci and the Gauss-Bonnet scalars, which are higher-order corrections on the curvature tensor added to the Einstein's gravitational action. Its goal is to give an alternative way to explain the evolution of the universe as we currently understand it, by providing other possible options than dark energy and the initial singularity,  both arising from the SCM. The presence of this term has also a mathematical basis and has been suggested as a feasible contribution by some string models.

In this section we present the general formalism of \frg modified gravity and work on an illustrative example for reconstructing it. We also discuss the necessary stability conditions for this type of gravity.

\subsection{Formalism}
First, we introduce the gravitational action for vacuum \frg gravity:
\beq
\mathcal{S}=\frac{1}{2\kappa}\int{d^4 x \sqrt{-g}F(R,\G)},
\eeq
where $\G$ is the Gauss-Bonnet invariant, defined as
\beq
\G \equiv R^2 -4R_{\alpha\beta}R^{\alpha\beta}+R_{\alpha\beta\rho\sigma}R^{\alpha\beta\rho\sigma},
\eeq
and $R$ is the Ricci scalar, which results from the contraction of the Ricci tensor,$R_{\alpha\beta}$,  $R_{\alpha\beta\rho\sigma}$ being the Riemann tensor.

In order to obtain the gravitational equations of motion, we have to vary the gravitational action with respect to the metric tensor $g_{\mu \nu}$. Considering the flat Friedmann-Lemaitre-Robertson-Walker (FLRW) metric,
\beq
ds^2=-dt^2+a^2(t)(dx^2+dy^2+dz^2),
\eeq
the gravitational equations become (see, for example, \cite{odintsov1})
\begin{align}
\label{friedman eq 2}
    \begin{split}
      2\dot{H}F_R+8H\dot{H}\dot{F}_\G=H\dot{F}_R -\ddot{F}_R+4H^3\dot{F}_\G-4H^2\ddot{F}_\G ,\\
    6H^2F_R+24H^3\dot{F}_\G=F_R R-F(R,\G)-6H\dot{F}_R+\G F_\G, 
    \end{split}
\end{align}
where $F_\G \equiv \frac{\partial F}{\partial \G} ~~ ;F_R \equiv \frac{\partial F}{\partial R}$,
and $\G$ and $R$, in the FLRW metric, are equal to 
\beq
\label{R def i g def}
R=6(2H^2 + \dot{H}) ~;~\G=24H^2(H^2 + \dot{H}).
\eeq

\subsection{Reconstructing F(R,\textbf{$\G$}) gravity }
\label{sec: reconstructing method}

In order to reconstruct the universe evolution under \frg, only approximate solutions can be obtained analytically, since the equations of motion are highly complicated and one must do some approximation. Therefore, in order to obtain the cosmic evolution, we are led to use the reconstruction technique developed in \cite{odintsov1}. We focus on a method which allows us to find the \frg function once the Hubble parameter is known.

Following the mentioned analysis, we express the Ricci and the Gauss-Bonet scalar of Eq.~(\ref{R def i g def}) in terms of the \textit{e}-folding number, $N=\ln (a/a_0)$:
\beq
R(N)=6(2H^2(N)+H(N)H'(N)),
\label{R n}
\eeq
\beq
\G(N)=24H^2(N)(H^2(N)+H(N)H'(N)),
\label{G n}
\eeq
where the prime indicates derivative with respect to $N$. 

Now, we assume a specific form for the functional of the Hubble parameter as, for example,
\beq
\label{hubble rate functional}
H^2(N)\equiv P(N).
\eeq
Then, from Eq.~(\ref{R n}) and Eq.~(\ref{G n}) we get
\beq
R=12P(N)+3P'(N) ; ~~\G=24P^2(N)+12P(N)P'(N).
\label{R i G p}
\eeq
Therefore, the second Friedmann equation (Eq.~(\ref{friedman eq 2})) can be written as
\begin{align}
\begin{split}
\label{eq: 2nd friedman eq N}
P(N)&(6F_R+24P(N)F_{\G\G}\G'(N))-F_R R\\
&+F(R,\G)+6P(N)F_{RR}R'(N)-\G(N)F_\G=0,
\end{split}
\end{align}
which is a second order differential equation, which we can solve. 

Hence, in order to obtain the solution for the \frg expression, we may first propose the Hubble rate we want to use, in order to relate it with $P(N)$, $R$ and $\G$ and, finally, solve the differential equation.

As an illustrative example of the reconstruction method described, we now propose the following expression for the Hubble rate:
\beq
\label{H proposal}
H^2(N)=P(N)=P_1 \exp{\beta N},
\eeq
with $P_1>0$. Then, by substituting Eq.~(\ref{H proposal}) into Eq.~(\ref{R i G p}), we get
\beq
N(\G)=\frac{1}{2\beta}\ln{\frac{\G}{12P_1^2(2+\beta)}}
\label{exemple N rec}
\eeq
\beq
P(\G)=\frac{\G^{1/2}}{(12P_1^2(2+\beta))^{1/2}}.
\label{exemple P rec}
\eeq

Moreover, in order to avoid very complicated equations, in what follows we will restrict to models with an \frg gravity expression of the form
\beq
\label{eq:type frg}
F(R,\G) = R + f(\G).
\eeq
Then, plugging Eq.~(\ref{exemple P rec}) into Eq.~(\ref{eq: 2nd friedman eq N}) and considering Eq.~(\ref{eq:type frg}), we find the differential equation
\beq
\label{fredmann dif eq reconstr}
\frac{2\beta \G^2}{(2+\beta)^2}(3+2\beta)F_{\G \G} + 3^{1/2}\frac{\G^{1/2}}{(2+\beta)^{1/2}}-\G F_\G + F(\G)=0.
\eeq

Quite often, the resulting differential equation may be very difficult to solve. If this happens, some leading order approximation will be needed. However, this is not  the current case, where Eq.~(\ref{fredmann dif eq reconstr}) can be solved in an analytic way, giving rise to
\beq
\label{G function reconstruct infl}
F(\G)=\frac{4b}{a-2}\G^{1/2}+F_0\G^{1/a}+F_1\G,
\eeq
with $F_0$ and $F_1$ being integration constants and
\beq
a=\frac{2\beta(3+2\beta)}{(2+\beta)^2}~~;~~b=\frac{3^{1/2}}{(2+\beta)^{1/2}}.
\eeq

As another example, we focus on the case where the scale factor has exponential form, which fits with the characteristics of the bouncing cosmologies discussed in Sect.~\ref{sec: intro}.

We consider the  scale factor
\beq
\label{eq: expo a}
a(t)=e^{\alpha t^2},
\eeq
which has a Hubble rate of the form
\beq
\label{eq: expo h}
H(t)=2\alpha t.
\eeq
As has been mentioned previously, the first step is to rewrite the variables we work with in terms of $N$, namely
\beq
\label{eq: expo h n}
H^2(N)\equiv P(N)= 2\alpha N.
\eeq
Since we are interested in discussing the results at late evolution stages, we perform the limit when $N\to \infty$, getting $H'\ll H$, what leads to a new expression for $\G(N)$, as
\beq
\G(N)\simeq 96 \alpha^2 N^2.
\eeq
Then, we obtain
\beq
\label{eq:N de g expo}
N(\G)=\frac{\G^{1/2}}{4\alpha\sqrt{6}},
\eeq
and, consequently
\beq
\label{eq: p de g expo}
P(\G)=\frac{\G^{1/2}}{2\sqrt{6}}.
\eeq
Furthermore, working with an \frg gravity of the type Eq.~(\ref{eq:type frg}) allows to get the second Friedmann equation, Eq.~(\ref{friedman eq 2}), as
\beq
\label{eq:friedman amb frg}
24(P(N))^2\G'(N)F_{\G\G}-\G(N)F_{\G}+6P(N)=0.
\eeq

At the end, we must solve the following non-trivial differential equation
\beq
\label{eq:expo differential}
\frac{48\alpha}{\sqrt{6}} \G^{3/2}F_{\G\G}-\G F_\G+F(\G)+\frac{\sqrt{6}}{2}\G^{1/2}=0.
\eeq
Fortunately, in \cite{odintsov1}, after an exhaustive analysis, the following approximate solution for this second order differential equation was found
\beq
\label{eq: fg in expo}
F(\G)=F_1 G^{0~0}_{1~1}(-\sqrt{\frac{\G}{6}}|0,2),
\eeq
with $F_1$ being an integration constant. Thus, the final expression for the \frg function reads
\beq
\label{eq: frg expo}
F(R,\G)= R+ F_1 G^{0~0}_{1~1}(-\sqrt{\frac{\G}{6}}|0,2).
\eeq
\subsection{Stability Conditions}
\label{sec:stability cond}
In this part, we introduce the necessary conditions for the background stability in the time evolution.

Following the procedure in \cite{escofet}, we rewrite Eq.~(\ref{eq: 2nd friedman eq N})  in terms of the functional $P(N)$, only, as follows
\begin{align}
    \label{friedman only pp}
    \begin{split}
24&P^2(N)(48P(N)P'(N)+12P(N)P''(N)\\
&+12P'^2(N))F_{\G\G}+6P(N)(12P'(N)+3P''(N))F_{RR}\\
&-(24P^2(N)+12P(N)P'(N))F_\G -(6P(N)+P'(N))F_R\\
&+ F(R,\G) = 0.
\end{split}
\end{align}
Taking now into account Eq.~(\ref{eq:type frg}) leads to $F_{RR}=0$ and $F_R=1$, and the previous expression becomes
\begin{align}
    \label{eq: friedman only p frg}
    \begin{split}
24&P^2(N)(48P(N)P'(N)+12P(N)P''(N)+12P'^2(N))F_{\G\G}\\
&-(24P^2(N)+12P(N)P'(N))F_\G +6P(N)+ F(\G) = 0,
\end{split}
\end{align}
where the functional $P(N)$ is actually the background solution. Thus, since we are interested in finding the stability of this solution, we must perform the variation over its background  $P(N)=P_0(N) + \delta P(N)$. In this way, we obtain
\beq
\label{eq: stability cond eq init}
\mathcal{J}_1 \delta P''(N) + \mathcal{J}_2 \delta, P'(N)+\mathcal{J}_3 \delta P(N)=0 ,
\eeq
with
\beq
\label{eq: j1}
\mathcal{J}_1 \equiv 288 P_0^3(N)F''(\G_0) ,
\eeq
\begin{align}
    \begin{split}
    \label{eq: j2}
\mathcal{J}_2=&432P_0^2(N)((2P_0(N)+P'_0(N))F''(\G_0)\\ &+8P_0(N)((P_0''(N))^2+P_0(N)(4P'_0(N)+P''_0(N))))   , 
    \end{split}
\end{align}
\begin{align}
    \begin{split}
    \label{eq: j3}
\mathcal{J}_3=&6(1+24P_0(N)((-8P_0^2(N)+3(P'_0(N))^2\\
&+P_0''(N))F''(\G_0)+24P_0(N)(4P_0(N)+P'_0(N))((P'_0)^2\\
&+P_0(N)(4P'_0(N)+P''_0(N)))F'''(\G_0)))  .
    \end{split}
\end{align}
Therefore, the stability conditions are $\frac{\mathcal{J}_2}{\mathcal{J}_1} > 0$ and $\frac{\mathcal{J}_3}{\mathcal{J}_1} >0$.

\section{Bounce in \frg Gravity}
\label{sec: bounce in frg}

In this section, we reconstruct the \frg gravity for a hyperbolic and for a matter-dominated bounce scenarios, and we discuss the stability of their cosmological solutions at late evolution stages.

\subsection{Hyperbolic Cosine Model}
\label{sec: hyper}

We first focus on the hyperbolic cosine model, which is characterized by the following scale factor
\beq
\label{eq: hyper a}
a(t)=\cosh(\lambda t),
\eeq
and, accordingly, by the corresponding Hubble parameter
\beq
\label{eq: hyper h}
H(t)=\lambda \tanh(\lambda t).
\eeq

In order to reconstruct the \frg gravity, we perform the same procedure as in Sect.~\ref{sec: reconstructing method}. First, we obtain the e-folding number and the function $P(N)$
\beq
\label{eq: hyper h n}
H^2(N)\equiv P(N)= \lambda^2 (1-e^{-2N}).
\eeq
Since we are interested in getting and discussing the solution at late time stages, we express $N(\G)$ and $P(\G)$ in the approximation $N\to \infty$, namely
\beq
\label{eq:N de g hyper}
N(\G)=\ln{\left(\frac{2\lambda\sqrt{6}}{\sqrt{24\lambda^2 - \G}}\right)},
\eeq
\beq
\label{eq: p de g hyper}
P(\G)\simeq\lambda^2.
\eeq
If we now insert Eq.~(\ref{eq:N de g hyper}) and  Eq.~(\ref{eq: p de g hyper}) into the Friedmann equation, Eq.~(\ref{eq:friedman amb frg}), we get the differential equation to solve
\beq
\label{eq: hyper diferential}
-\G F_{\G\G} + F(\G) +\lambda^2 \G=0,
\eeq
which leads to
\beq
\label{eq: fg hyper}
F(\G)=\lambda^2 \G \ln{(\G)} + K\G,
\eeq
being $K$ an integration constant.

Hence, our reconstructed \frg gravity which achieves Eq.~(\ref{eq: hyper a}) bounce behavior is
\beq
\label{frg hyper}
F(R,\G)= R + \lambda^2 \G \ln{(\G)} + K\G.
\eeq

Now we proceed to discuss the stability of this cosmological model at late stages.
Taking into account that Eq.(~\ref{eq: hyper h n}) at a late phase becomes $P(N)\simeq \lambda^2$, its derivatives are $P'(N)\simeq0$ and $P''(N)\simeq0$.
Consequently, we search for the expressions for the stability conditions by first obtaining $\mathcal{J}_1$, $\mathcal{J}_2$ and $\mathcal{J}_3$, as
\beq
\label{eq: j1 hyper}
\mathcal{J}_1=288\lambda^6 F''(\G_0),
\eeq
\beq
\label{eq: j2 hyper}
\mathcal{J}_2\simeq432\lambda^4(12\lambda^6 F''(\G_0)=864 \lambda^6 F''(N),
\eeq
\beq
\label{eq: j3 hyper}
\mathcal{J}_3\simeq 6(1-192\lambda^6 F''(\G_0)).
\eeq
At this moment, we are able to discuss whether the stability conditions are satisfied in this model or not:
\beq
\label{eq: stab cond hyper 1}
\frac{\mathcal{J}_2}{\mathcal{J}_1}=3>0,
\eeq
\beq
\label{eq: stab cond hyper 2}
\frac{\mathcal{J}_3}{\mathcal{J}_1}=\frac{1}{48\lambda^6F''(\G_0)}-4>0
\eeq

As we can see, this hyperbolic model fulfills the stability conditions at late stages.

\subsection{Matter-Dominated Bounce Model}
\label{sec: matter bounce}
In this section we focus on a possible matter-bounce scenario \citep{saridakis}, which is characterized by the following scale factor
\beq
\label{eq:matter bounce a}
a(t)=a_b(1+qt^2)^{1/3},
\eeq
where $a_b$ is the scale factor at the bounce and $q$ is a constant. The corresponding Hubble parameter is
\beq
\label{eq:matter bounce h}
H(t)=\frac{2qt}{3(1+qt^2)}.
\eeq

Following the same reconstruction method as in the previous sections, we first obtain $P(N)$
\beq
\label{eq:matter bounce h n}
H^2(N)\equiv P(N)= \frac{4}{3}e^{-3N},
\eeq
where $N\simeq \frac{2}{3} \ln{(t)}$. Then,  putting Eq.~(\ref{eq:matter bounce h n}) into Eq.~(\ref{R i G p}), we get the  number of e-folds in terms of the Gauss-Bonnet scalar, $N(\G)$,  and also $P(\G)$, as
\beq
\label{eq:N de g mater}
N(\G)=\frac{1}{6}\ln{\left(-\frac{64}{27\G}\right)},
\eeq
\beq
\label{eq: p de g mater}
P(\G)=\frac{\sqrt{-27\G}}{18}.
\eeq
Substituting Eq.~(\ref{eq: p de g mater}) and Eq.~(\ref{eq:N de g mater}) into the Friedmann equation, we are able to solve the differential equation, with the result
\beq
\label{eq:f(g) matter}
f(\G)= C_1\G^{1/12}+C_2 \G +\frac{2\sqrt{3}}{5}\sqrt{-\G},
\eeq
and, therefore
\beq
\label{eq:frg matter}
    F(R,\G) = R+ C_1\G^{1/12}+C_2 \G +\frac{2\sqrt{3}}{5}\sqrt{-\G}.
\eeq

As in the previous section, we are interested in discussing if this particular matter-bounce scenario fulfills the stability conditions. Taking into account Eq.~(\ref{eq:matter bounce h n}), the derivatives of $P(N)$ are 
\beq
\label{eq: p prima matter}
P'(N)= -\frac{4}{3} e^{-3N},
\eeq
\beq
\label{eq: p prima prima matter}
P''(N)= 4 e^{-3N}.
\eeq
Therefore, considering the same leading order approximation at late stages, $N\to \infty$, we obtain the expressions for $\mathcal{J}_1$, $\mathcal{J}_2$ and $\mathcal{J}_3$:
\beq
\label{eq: j1 matter}
\mathcal{J}_1=\frac{1048}{81}e^{-9N} F''(\G_0),
\eeq
\begin{align}
\begin{split}
\label{eq: j2 matter}
\mathcal{J}_2&=\frac{256}{3}e^{-6N}(-\frac{4}{9}e^{-3N}F''(\G_0)+\frac{13312}{243}e^{-9N})\\
&\simeq-\frac{1024}{27}e^{-9N}F''(\G_0)
\end{split}
\end{align}
and 
\beq
\label{eq: j3 matter}
\mathcal{J}_3\simeq 6.
\eeq
Finally, we arrive the following expressions
\beq
\label{eq: stab cond matter 1}
\frac{\mathcal{J}_2}{\mathcal{J}_1}=-\frac{2}{3}<0,
\eeq
\beq
\label{eq: stab cond matter 2}
\frac{\mathcal{J}_3}{\mathcal{J}_1}=\frac{486}{2048}\frac{e^{9N}}{F''(\G_0)}=\infty>0.
\eeq

As we can observe from these relations, this specific scenario does not accomplish the background stability conditions at late time evolution.

\section{Conclusions}
\label{sec: conclusions}
In this paper, after discussing the main characteristics of \frg modified gravity, we have presented in detail the reconstruction method used in the rest of the paper and considered some illustrative examples, in particular the exponential bounce model. We have also described the conditions needed for a cosmological model evolution to be stable and subsequently applied the reconstruction method for the hyperbolic cosine bounce model, in which case we have successfully found a possible \frg expression which achieves the goal, and we have then proven its stability at late cosmological times (for interesting alternative methods, see \citep{Capozziello_conclusions1,Capozzielo_conclusions2,2009PhLB..Capozzielo_conclusions3}). That this is rather non-trivial is proven by the second example, where we have carried out the same analysis in a matter-bounce cosmology, which is actually consistent with astronomical observations of the early universe, as discussed in Sect.~\ref{sec: intro}. Actually, we have proven that, at a later epoch, $N\to \infty$, this second specific model does not fulfill the stability conditions and, therefore, it is non-viable. This different stability behavior could be explained by the different evolution of the Hubble radius in both cases and, therefore, a different generation era of the perturbation modes. In the cosine hyperbolic case it goes to zero asymptotically and therefore the perturbation modes generate near the bounce. On the other hand, in the matter-dominated bounce, the Hubble radius diverges asymptotically and thus the
perturbation modes generate deeply in the contracting regime far away from
the bounce, suggesting that it may be inter-related with the results obtained in Sect. \ref{sec: bounce in frg}. In order to recover the stability in the matter bounce model, one could use the lagrange multyplier \frg model, which seems to make the matter bounce stable \cite{elizalde_tanmuy}.

\section*{Acknowledgements}
E.E. has been partly supported by MINECO (Spain), project FIS2016-76363-P, and by AGAUR (Catalan Government), project 2017-SGR-247. We thank Tanmoy Paul for interesting comments.



\nocite{*}

\bibliography{main}

\end{document}